\documentstyle[12pt,epsf]{article}

\newcommand {\bea}{\begin{eqnarray}}
\newcommand {\eea}{\end{eqnarray}}
\newcommand {\be}{\begin{equation}}
\newcommand {\ee}{\end{equation}}

\begin{document}


\title{
\begin{flushright}
\begin{small}
HUTP-97/A095, UCSB-97-24 \\
UPR-778-T, hep-th/9712077\\
\end{small}
\end{flushright}
\vspace{1.cm}
Gauge Theory, Geometry and the Large N Limit
}
\author{Vijay Balasubramanian,$^{(1)}$\thanks{vijayb@pauli.harvard.edu}
Rajesh Gopakumar$^{(1,2)}$\thanks{gopakumr@pauli.harvard.edu}
and Finn Larsen$^{(3)}$\thanks{larsen@cvetic.hep.upenn.edu}\\
\small (1) Lyman Laboratory of Physics, 
Harvard University, Cambridge, MA 02138\\
\small (2) Physics Department, Univ. of California at Santa Barbara,
Santa Barbara, CA 93106 \\
\small (3) David Rittenhouse Laboratories, University of Pennsylvania,
Philadelphia, PA 19104
}

\date{ }
\maketitle

\begin{abstract}
We study the relationship between M theory on a nearly lightlike
circle and $U(N)$ gauge theory in $p+1$ dimensions.  We define large
$N$ limits of these theories in which low energy supergravity is
valid. The regularity of these limits implies an infinite series of
nonrenormalization theorems for the gauge theory effective action, and
the leading large $N$ terms sum to a Born-Infeld form. Compatibility
of two different large $N$ limits that describe the same
decompactified M theory leads to a conjecture for a relation between
two limits of string theories.
 
\end{abstract}


\section{Introduction}
\label{sec:intro}

A common theme of recent work in string theory is that conventional
geometry should sometimes emerge as an effective low-energy
description from an auxiliary gauge theory. This surprising idea
emerges in its simplest form from an examination of the low-energy dynamics
of parallel D-branes~\cite{witten95a,dkps}. In this context, the
classical geometry of spacetime appears via dynamics on the quantum
moduli space of gauge configurations. The connection between gauge
theory and geometry has been exploited in M(atrix) theory, the
conjectured nonperturbative definition of M-theory~\cite{bfss,dlcq},
as well as in the related analysis of M(atrix)
strings~\cite{matstring}.  In these developments, the implied logic is
that gauge theory is simpler than string theory, and so constitutes a
preferred starting point for the discussion. In the present paper, we
pursue the reverse strategy: we consider limits of M-theory where
spacetime geometry can be reliably studied in supergravity, and use
these to infer properties of gauge theory effective actions.

Our starting point is the recent argument by Seiberg that relates M-theory 
on a nearly lightlike circle to the $U(N)$ gauge theory dynamics of $N$ 
D-branes~\cite{seibmat} (also see~\cite{senmat}). We consider several 
large $N$ limits where M-theory is adequately described 
by its low-energy supergravity approximation. We argue that these limits
are nonsingular and use the requirement that the corresponding gauge theory 
limits also be regular to find an infinite series of non-renormalization 
theorems for the loop expansion of the gauge theory effective action.

Several subtleties arise. For example, a null circle has zero proper
length; so low-energy M-theory on a nearly lightlike circle is in
danger of breaking down due to short distance effects. However, we
find that a large momentum along the circle exerts a pressure that
makes it spacelike and large. In our large $N$ limits the M-theory is
therefore decompactified in the bulk of spacetime and neither the
curvature invariants, nor the proper distance around the circle,
approach the 11 dimensional Planck scale. It would be a surprise if 11
dimensional supergravity could not be applied in such regimes where it
shows no sign of breaking down.

An important result for the gauge theory is that the leading large N
terms to all loops in the gauge theory effective description of
graviton scattering sum up to a Born-Infeld form. This result is
already known as a prediction of
M(atrix)-theory~\cite{bbpt,chepact,peract,minic,dealwis}, but we
recover it using different assumptions.  Our large $N$ limits are not
the 't Hooft scaling limits which usually make sense for a generic
gauge theory. We will comment on how our alternative limits seem to
make sense as a consequence of the non-renormalization theorems.

Additionally, we identify certain nonplanar diagrams in the gauge
theory with corrections to the 11 dimensional low-energy effective
action that are suppressed by powers of the Planck length. We also
present a conjecture arising from a comparison between two different
large $N$ limits which should, at least naively, yield the same
decompactified M theory. This conjecture essentially says that string
theory in the large $N$ gauge theory limit is equivalent to the same
string theory, but in the rather different domain specified in 
Sec.~\ref{sec:twolim}.
This equivalence is in a similar spirit (but not exactly the same) as
the expected simultaneous validity of supergravity and large $N$ gauge
theory in the near horizon limit of black holes~\cite{dps}.


Sec.~\ref{sec:DLCQ} summarizes the argument that
relates M-theory on a nearly lightlike circle and a transverse $T^p$, to
$U(N)$ gauge theory in $p+1$ dimensions~\cite{senmat,seibmat}
and discusses graviton scattering in terms of effective
actions in gauge theory and supergravity.  In Sec.~\ref{sec:low} we specify
the large $N$ limits that allow descriptions in both gravity and gauge
theory. The discussion for gravitons is generalized to extended objects 
and we give some further arguments for the Born-Infeld resummation of the
leading contributions to graviton scattering.

\section{M-theory and Gauge Theory}
\label{sec:DLCQ} 

\subsection{Preliminaries}
Consider M-theory on a nearly lightlike circle defined by 
the identifications
$(x_{11},t) \sim ( x_{11}+\sqrt{R^2/2 + R_S^2}, t-R/\sqrt{2})$,
where $R_S\ll R$.  Also introduce a transverse $T^p$ specified by
$r^i \equiv r^i + R^i$.\footnote{We use $R^i$ for the {\it
circumference} of a circle, following many recent works.} Working in
this nearly lightlike frame we consider a probe graviton with
longitudinal momentum $P_p^+ = K/R$ impinging on a target graviton
with $P_t^+ = N/R$.  We will always be interested in situations with
$K \ll N$.  The probe has small transverse velocity $v$ in a
noncompact direction, in the center of mass frame where the heavy
target is nearly stationary.  The probe and target
are separated by $b$ in a noncompact direction other than the direction 
of motion.

Seiberg's kinematic analysis~\cite{seibmat} exploits 11 dimensional 
Lorentz invariance to apply a boost of rapidity $\alpha$ to the nearly 
lightlike circle where $\sinh \alpha = R/\sqrt{2}R_S$.  
Winding modes are important after the boost and
can be accounted for by T-dualizing the torus to get a theory of K 
Dp-brane probes interacting at low energies with a target of N Dp-branes.  
After further uniform rescaling of all lengths as $\tilde{l}= l (R_S/R)$
the M-theory scattering problem reduces
to the dynamics of Dp-branes with effective string coupling 
and string scale:
\begin{equation}
\tilde{g}_S = {(R_S M_P)^{3/2} \over V_T (R_S M_P^3)^{p/2} }  
~~~~~~~~;~~~~~~~~ \tilde{M}_S^2 = {R^2 M_P^3 \over R_S}
\end{equation}
where $V_T = \prod_i R_i $ is the volume of the original transverse
torus in M-theory.  The probe branes have a velocity $\tilde{v} = v
R_S/R$, kinetic energy $E = Kv^2/R$, and are separated from the target
branes by  $\tilde{b} = b R_S/R$ where $v$ and $b$ are
the original 11 dimensional velocity and separation.  The 
T-dual transverse torus has cycles of  sizes $\Sigma_i = 1/R_i R M_P^3$.

With these parameters, for $p\leq 3$, the $R_S \rightarrow 0$ limit
yields the standard gauge theory description of the low-energy,
short-distance dynamics of Dp-branes.  The action for the $K$ probes
and $N$ target branes is the dimensional reduction of minimally
supersymmetric $U(N+K)$ Yang-Mills in 10 dimensions to $(p+1)$
dimensions:
\begin{equation}
S_0 = {1 \over g^2_{YM}} \int d^{p+1}x \, {1 \over 4} {\rm Tr}(F^2)
\end{equation}
To perform the dimensional reduction, for $i>p$ the gauge fields $A^i$
are interpreted as scalar fields $X^i$ and all derivatives
$\partial_i$ vanish.  We will relate the fields $A$ and $X$ to the
parameters of the scattering branes in the next section.  The coupling
$g^2_{YM}$ (finite as $R_S \rightarrow 0$ ) is given by:
\begin{equation}
g^2_{YM} = {\tilde{g}_S \over \tilde{M}_S^{p-3}} =
           {(R \, M_P^2)^3 \over V_T \, (R \, M_P^3)^p}
\label{eq:gYM}
\end{equation}
For $p>3$, the $R_S \rightarrow 0$ limit is more
complicated because it leads to large string coupling.

\subsection{Scattering in gauge theory}
\label{sec:scattg}
The separation of probe from target in M-theory is $b$ and, after the above
rescaling of units, the separation between $Dp$-branes is $\tilde{b} =
b R_S/R$ in the boosted theory.  The positions of branes in the gauge
theory description are given by eigenvalues of the adjoint scalars
$X^i$ and the $K$ probes can be separated from the $N$ targets by a
VEV:
\begin{equation}
X^i = \pmatrix{0_{N\times N} & 0 \cr 0 &  x \, I_{K \times K}}
\label{eq:higgs}
\end{equation}
We can relate $x$ to the physical
separation $\tilde{b}$ as $x = \tilde{b} \tilde{M}_S^2 = b~R M_P^3$.
The M-theory probe moves at a transverse velocity of $v$ and the 
boost that
produces the related system of Dp-branes rescales the velocity to
$\tilde{v} = v R_S/R$.  This velocity appears as the only nonvanishing
component of the electric field in the gauge theory:
\begin{equation}
F_{0j} = \partial_0 X_j = \pmatrix{0_{N\times N} & 0 \cr
                                   0 & f  \, I_{K \times K}}  
\label{eq:electric}
\end{equation}
Here $f$ is related to the physical velocity of the probe as $f =
\tilde{v} \tilde{M}_S^2 = v~R M_P^3 $.
The VEV in Eq.~\ref{eq:higgs} spontaneously
breaks the gauge group $U(N+K)\rightarrow U(N)\times U(K)$ and
exchange of the resulting W-bosons produces the interaction.  This can
be studied efficiently by constructing the effective action as a
function of $b$ and $v$ that results upon integrating out the W-boson.

The general form of the perturbative effective action resulting from
this  procedure can be understood using the dimensional analyses
of~\cite{ggr1, bbpt}:
\begin{equation}
L_{\rm eff} = {f^2 \over 2\, g^2_{YM}} \left[1 + \sum_{L,m = 1}^\infty c_{Lm}
                \left({g^2_{YM}f^2\over x^{7-p}}\right)^L~
                  \left({f^2\over x^4}\right)^{m-L} \right]
\end{equation}
Here $L$ counts the number of Yang-Mills loops and $m+1$ counts the
insertions of $f^2$.  
In M theory variables $x = b~R M_P^3$, $f = v~R
M_P^3$ and $g^2_{YM}$ is given in Eq.~\ref{eq:gYM}.  
At $L$ loops, planar diagrams will have a total power of $N$ and $K$ adding 
to $L+1$.  Nonplanar diagrams lower the total powers of $N$ and $K$ by an
even number.\footnote{The total power of $N$ and $K$ is reduced by an
even number because $U(N)$ diagrams in 't Hooft double line notation
are orientable.} So at $L$ loops the general $N$
and $K$ dependence is $N^{L-2n-q} K^{1 +q}$ where $2n$ counts the
degree of nonplanarity, and $q+1$ counts how many boundaries lie on
the probe in a 't Hooft double line representation of the gauge theory
diagram. The general form of the effective probe Lagrangian becomes:
\begin{eqnarray}
L_{\rm eff} &=& 
{K \, v^2 \over 2\, R}~\left[1+\sum_{L,m = 1}^\infty 
\sum_{n=0}^{\lfloor (L-1)/2 \rfloor }
\sum_{q=0}^{L-1-2n} \, d_{Lm}(n,q)~K^{q}~N^{2n-q} \right. \nonumber \\
&\times & \left. \left({N\,v^2\over R^2 M^9_p V_T~b^{7-p}}\right)^{L}~
\left(v^2 \over R^2 M^6_p b^4\right)^{m-L}    \right]
\label{eq:general}
\end{eqnarray}
(We have integrated this effective Lagrangian over the $p$ spatial
dimensions of the probe cancelling a factor of the dual torus volume.)
The tree level ($v^2$) term is proportional to $K$ because there are
$K$ probes and since the VEVs we have chosen are symmetric between the
probes.  Similarly, the factors of $N$ arise because the probes
interact with each brane in the target symmetrically.  In principle,
each factor of $K$ and $N$ should be replaced by traces over $U(K)$
and $U(N)$ gauge indices respectively and the effective action should
be written covariantly in terms of $F$.  

Eq.~\ref{eq:general} is an analysis of  
the Feynman loop expansion of the gauge theory.
However, there can also be contributions to the coefficients
that are not 
captured by Feynman diagrams. Nonperturbative effects can also produce new 
terms that are simply not seen in perturbation theory. Indeed, we certainly 
know that they are necessary
in considering graviton scattering in the presence of transverse p-tori for
$p \geq 2$~\cite{polchpoul,bfss2}.  Also, we have not taken account of the
elusive bound-state wavefunction governing the interacting clumps of branes
in the gauge theory.  Such bound state effects can easily add corrections
to Eq.~\ref{eq:general} with modified dependences on $N$ and $K$.  Analysis
of such effects is very difficult, and so we are trying to learn as much as 
possible from general principles.


\subsection{Scattering in supergravity}
\label{sec:sugrascatter}

The limit where the scattering is described by gauge theory can in some
cases also be treated accurately in supergravity. To determine when this
is the case consider 11 dimensional supergravity compactified on a {\em
spatial} circle of length $R_S$ and a transverse $T^p$.  This theory
has plane wave solutions carrying $N$ units of momentum in the compact
direction:
\begin{eqnarray}
ds^2 &=& dx^+ dx^- + dx_1^2 + \cdots + dx_9^2 + \tilde{D} (dx^+)^2 
\label{eq:10met}\\
\tilde{D} &=&  { c_p \,N \over (r M_P)^{7-p} (R_S M_P)^2 (V_T
                     M_P^p)}  
\label{eq:10D}\\
c_p &=& 4\pi^{p-1\over 2} \Gamma\left({7-p \over 2}\right) 
\label{eq:c}
\end{eqnarray}
where $x^\pm = x^{11} \pm t$.  Supergravity becomes inaccurate below the
string scale so it may appear that this solution is only valid for $ r \gg
(R_S M_P^3)^{-1/2}$. This would be a very severe requirement because we
would like to take $R_S \rightarrow 0$. Happily this conclusion is too
hasty: Eq.~\ref{eq:10met} represents a solution on a spacelike circle
$x_{11} \equiv x_{11} + R_S$.  This metric can be rewritten as:
\begin{eqnarray}
ds^2 &=& (-dt^2 + dx_{11}^2 + dx_1^2 + \cdots + dx_9^2) + 
\tilde{D} (dx_{11} + dt)^2 \label{eq:10met2} \\
\tilde{D} &=&  { c_p \,N \over (r M_P)^{7-p} (R_S M_P)^2 (V_T
                     M_P^p)}  
\end{eqnarray}
When the 11th circle is of length $R_S$ we only expect supergravity to be valid
at distances greater than the string length:  $ r \gg (R_S M_P^3)^{-1/2}$.  
However, Eq.~\ref{eq:10met2} shows that the physical length squared of the 
circle is:
\begin{equation}
l^2 = \left(1 + { c_p \,N \over (r M_P)^{7-p} (R_S M_P)^2 (V_T
                     M_P^p)} \right) R_S^2 
= R_S^2 +  { c_p \,N \over (r M_P)^{7-p} M_P^2 (V_T
                     M_P^p)}
\end{equation}
So, {\em independently} of $R_S$, in the large $N$ limit (which we
will take in the next section) the circle is much bigger than the 11
dimensional Planck length for any $r < \infty$.  What is more, it can
be checked that the curvature invariants scale inversely with $N$ and
so remain small in the limit.  This strongly suggests that, for large
$N$, the solution remains valid even in the $R_S \rightarrow 0$ limit.
Essentially, the momentum along the circle exerts a pressure that
decompactifies the solution allowing a valid treatment in 11
dimensional supergravity.  From the 10 dimensional perspective, the
solutions discussed above are D-particles or extremal black holes.  We
find that in the large $N$ limit the throat region of the black hole 
extends far from the source and can be treated reliably in supergravity. 
Essentially, the local dilaton becomes large and so the solution should
be considered 11 dimensional.

    Now we use 11 dimensional Lorentz invariance to boost the compact
direction by a rapidity $\beta$ with $\sinh \beta = -R/\sqrt{2} R_S$.  This
converts the spacelike circle of length $x_{11} \equiv x_{11} + R_S$ into a
nearly lightlike circle  and the boosted metric (to
leading order in $R_S/R$) is:
\begin{eqnarray}
ds^2 &=& dx^+ dx^- + dx_1^2 + \cdots +dx_9^2  + D (dx^+)^2 
\label{eq:boostmet} \\
D &=&  { c_p \, N \over (r M_P)^{7-p} (R M_P)^2 (V_T M_P^p)}
\label{eq:boostharm}
\end{eqnarray}
Since $R_S$ has completely dropped out of the leading order boosted metric,
{\it we can take $R_S$ to zero freely without affecting the validity of the
solution}.  In other words, the boosted solution is valid for all $r \gg
1/M_P$.  This is surprising because, as emphasized
in~\cite{seibmat,heller}, the nearly lightlike compactified theory is
related by a boost to the theory on a small spacelike circle where the
string length would seem to be the relevant scale. In fact, as emphasized
above, the supergravity solution on a spacelike circle has a much larger
regime of validity than naively expected.  A physical reason for the
validity of the boosted solution at short distances is that $x^+$, the
putative compact ``null'' direction, is in fact {\em spacelike} for $r <
\infty$.  What is more, in the large $N$ limit (which we will take in the
next section), this {\em spacelike} circle is much larger than $1/M_P$ for
any finite $r$.  This strongly suggests that, for large $N$, we should be
able to do reliable 11 dimensional supergravity calculations in this
background.

In the limit of small $R_S$, the effective Lagrangian for a probe graviton
carrying longitudinal momentum $P^+ = K/R$ has been constructed
in~\cite{bbpt}:
\begin{equation}
L_g = - {K\over R} D^{-1} (\sqrt{1 - D v^2} - 1)
\label{eq:grav}
\end{equation}

\section{Low energy, Large N limits }
\label{sec:low}

In this section we are going to take $N \rightarrow \infty$ limits
while $R \rightarrow \infty$ in such a way that all energies and
momenta are small.  
It is important that we take $ N$ and $ R$ to
$\infty$ simultaneously and in a correlated fashion since we can
always rescale $R$ itself by a boost.  We will consider two kinds of
limits.  First, following~\cite{seibmat}, we take $R_S \rightarrow 0$
before taking $R$ and $N$ to infinity.  
From the discussion of supergravity solutions in the previous section 
we find that this order of limits has a low energy supergravity 
description. This result allows us to sum the leading $N$ terms in
the gauge theory effective action to all loops.  Finally we consider a
second limit in which $R_S$ is held finite while $N$ and $R$ are taken
to infinity.  This is another decompactification limit of M theory.
Assuming that the decompactification limit is unique leads to a
specific conjecture for a relation between long and short distance
string theory.


\subsection{A supergravity limit (L1)}
\label{sec:l1}
We begin by taking the $R_S \rightarrow 0$ limit of the nearly
lightlike circle.  Then we perform the following
substitution on the parameters of the M theory:
\begin{equation}
L1:~~~~~~~
R \rightarrow  \lambda R_0 ~~~~~;~~~~~
N \rightarrow \lambda^2 N_0 ~~~~~;~~~~~
K \rightarrow \lambda^2 K_0
\label{eq:l1limit} \\
\end{equation}
and take the scaling limit $\lambda\rightarrow\infty$. We will begin by 
arguing that this limit should be nonsingular and well described 
by the low-energy effective action of M theory.

Since $N$ and $K$ are getting very large there is some danger that
$L1$ is a singular limit since the energies of the probe and target
are very large.  There are two energies relevant to our scattering
processes: the energy stored in the longitudinal momentum which acts
like a ``mass'', and the energy in the transverse motion that acts
like a ``kinetic'' energy.  We start with the ``kinetic'' energy in
the transverse motion of the probe.  We can study this energy in a
boost invariant way by examining the transverse momentum:
\begin{equation}
P_\perp^2 = P^+ P^- \sim {K^2 v^2 \over R^2}
\end{equation}
In the decompactification limit, we expect the interactions between probe
and target to be under control when the momentum density is small.  The
physical length of the compact direction scales proportionally to $R$, and
so an appropriate requirement is:
\begin{equation}
{P_\perp \over R} \sim {K v \over R^2} \ll M_P^2
\label{eq:enercond}
\end{equation}
This relation is {\em independent} of the scaling limit $L1$ and
simply requires a suitably small velocity.

Next we may worry that the mere fact that the longitudinal momenta of
both probe and target ($P_p^+ = K/R$ and $P_t^+ = N/R$) are diverging
will imply that a low energy description will break down due to strong
gravitational effects.  But we have already argued that the
supergravity solution in Eq.~\ref{eq:boostmet} remains valid at small
$R_S$.  Evidence against the breakdown of this solution is provided by
the smooth limit of the harmonic function $D$ under the $L1$ scaling.
So there is no reason to expect strong gravitational
effects from the increasing ``mass'', essentially because the ``mass''
density is not divergent in the decompactified $L1$ limit.
Furthermore, as $N\rightarrow \infty$, we have argued that the compact
direction becomes spacelike and large while curvature invariants
remain small.  This too suggests that gravitational effects are small.

Quantum corrections are immediately recognized as benign from the 11
dimensional point of view because all length scales are large.  From the 10
dimensional point of view the local value of the dilaton is large close to
the branes, suggesting a strong coupling region in the interaction of
D-branes and hence the gauge theory.  However, as explained in~\cite{dkps},
this apparent divergence is taken care of correctly by the gauge theory so
that it is in fact the asymptotic dilaton that measures the magnitude of
quantum corrections.  The reason is that, at short distance, handles on
diagrams with several boundaries are suppressed. In this way, it is
precisely the circle of vanishing size at infinity, which is problematic
for the 11 dimensional supergravity, that keeps the 10 dimensional theory
under control.

Finally, consider the effective action for a supergravity probe given in
Eq.~\ref{eq:grav}.  Expanding in powers of $v$ gives a leading term
$Kv^2/2R$.  (Note that this the same as the tree level term in the gauge
theory effective action in Eq.~\ref{eq:general}.)  In the $L1$ scaling
limit, this ``kinetic energy'' term grows linearly in $\lambda$.  This is a
simple reflection of the fact that the total probe longitudinal momentum,
which acts as an effective ``mass'', grows linearly in the $L1$ limit.  The
growth with $\lambda$ is just a consequence of the extensivity of the probe
effective action and can be divided out.  The condition that permits
expansion of the supergravity effective action (Eq.~\ref{eq:grav}) in
powers of $v$ is:
\begin{equation}
D v^2 = {c_p N v^2 \over (r M_p)^{7-p} (R M_P)^2 (V_T M_P^p) } \ll 1
\label{eq:graveff}
\end{equation}
Since $D$ has a well defined limit under the $L1$ scaling, this requires small
velocities and/or large distances in the scattering process.

Since we began by taking $R_S \rightarrow 0$, the discussion in
Sec.~\ref{sec:DLCQ} applies and M theory can be studied in the gauge
theory description of low energy Dp-branes.  So, to analyze the effect
of the $L1$ limit on graviton scattering in M theory, we should study
the effect it has on the gauge theory dynamics.  Applying the scaling
to the effective action in Eq.~\ref{eq:general} we find that the tree
level term scales linearly as $\lambda$.  As discussed above, this
overall scaling arises simply because of the extensivity of the probe
action.  After dividing out this overall scaling with $\lambda$, the
individual terms in the rest of the Lagrangian scale as $\lambda^{2(L
- 2n - m)}$.  So as $\lambda \rightarrow \infty$, the terms with $2n +
m > L$ vanish, while the terms with $2n + m < L$ diverge.  However, as
discussed above, there are good reasons to believe that the $L1$ limit
is a nonsingular limit of M theory.  If so, we must conclude that the
coefficients $d_{Lm}(n,q)$ vanish for $2n + m < L$.  For example, for
$n=0$ we are dealing with planar diagrams, and we are finding that all
$L$ loop planar diagrams with less than $2L+2$ velocity insertions
must vanish.  In other words, by assuming 11 dimensional Lorentz
invariance and the plausible existence of the limit $L1$, we have
shown an infinite number of non-renomalization theorems for the gauge
theory effective action.  For $p=0$, one of these theorems says that
the one-loop $v^4$ term is perturbatively exact in the large N loop
expansion.  In fact, in this case, the vanishing of the coefficent of
the two loop $v^4$ term ($L=2, m=1$) was checked in ~\cite{beckers1}.
As we mentioned at the outset, our claims strictly concern terms of a
form that can appear in the loop expansion.  It is known, for example,
that for $p=2$ there are non-perturbative contributions to the $v^4$
term ~\cite{dinseib} but these do not take the form of the terms in
the loop expansion of Eq.~\ref{eq:general}.

The finite terms in the $L1$ limit are those that have $L=2n+m$. We insert
this value of $L$ and then relabel the parameter $m$ as $L$.  After an
overall rescaling by $1/\lambda$ we find:
\begin{equation}
L_{L1} = 
{K \, v^2 \over 2\, R} ~\left[1+
\sum_{L = 1}^\infty \sum_{q=0}^{L-1}
\, d_{LL}(0,q)~K^{q}~N^{-q} 
\left({N\,v^2\over R^2 M^9_p V_T~b^{7-p}}\right)^{L}~(1+S_{L1})
\right]
\label{eq:L1action}
\end{equation}
Here $S_{L1}$ is a series of corrections suppressed by inverse powers
of $M_P^6$:
\begin{equation}
S_{L1} = \sum_{n=1}^\infty f_{Lq}(n) \,
\left({v^2\over R^2 M^9_p V_T~b^{7-p}}\right)^{2n}~ 
\left(v^2 \over R^2 M^6_p b^4\right)^{-2n}    
=\sum_{n=1}^\infty f_{Lq}(n) \, 
\left({1 \over M_P^3 V_T~b^{3-p} }\right)^{2n}
\end{equation} 
where $f_{Lq}(n)$ are constant coefficients.  We want to compare this
limiting action to low energy, decompactified M-theory.  Since $M_P$ is
finite in our limit, we expect corrections to scattering amplitudes that
arise from subleading corrections to the M theory low energy effective
action.  Russo and Tseytlin have shown that the curvature corrections to
M-theory that derive from perturbative corrections in string theory are of
the form ${\cal R}^{3k+1}/M_P^{6k-9}$~\cite{rtR4} where ${\cal R}$ is the
Riemann tensor.  In our limit of the gauge theory, the subleading terms in
$M_P$ are in the series $S_{L1}$ which collects terms arising from
nonplanar diagrams.  It is pleasing that $S_{L1}$ is a series in $1/M_P^6$
in accord with our expectation from Russo and Tseytlin.  The terms in the
effective action that have $n=0$ (and come from planar diagrams) should be
identified with the leading term in the M theory low energy effective
action - 11 dimensional supergravity. It is interesting that the quantum
corrections in M-theory should appear in ``diagonal'' bands ($L=2n+m$)
whereas the quantum expansion in the gauge theory is ``vertical'' (in $L$).

To study the planar diagrams with $n=0$ note that the effective loop
expansion parameter in Eq.~\ref{eq:L1action} is:
\begin{equation}
g^2_{\rm eff} = {N v^2 \over (b M_p)^{7-p} (R M_P)^2 (V_T M_P^p)}
\label{eq:gaugeeff}
\end{equation}
The condition that this should be small, so that the perturbative
treatment is valid, is precisely the same as the condition in
Eq.~\ref{eq:graveff} that permitted expansion of the supergravity
effective action in Eq.~\ref{eq:grav} in an eikonal type series in
powers of velocity.  For $n=0$, the finite terms in the $L1$ limit
are, up to coefficients, an expansion in $Dv^2$ where $D$ is the
harmonic function in the supergravity solution of
Eq.~\ref{eq:boostmet}.  This means that the $n=0$ terms
in the gauge theory action that scale as $N^L K$ coincide (up to the
undetermined coefficients $d_{LL}(0,0)$) with the expansion in in
powers of velocity of the supergravity effective action in
Eq.~\ref{eq:grav}.  In the next section we will take a different large
$N$ limit that isolates just these terms.  In the $L1$ limit, the
total longitudinal momentum of the probe diverges and so back-reaction
effects are important.  The terms with higher powers of $K$ can be
interpreted as arising from back-reaction of the probe gravitons.  The
supergravity probe action in Eq.~\ref{eq:grav} was derived in the
absence of back-reaction - hence we cannot hope to compare it
directly to the gauge theory in the $L1$ limit.

\subsection{A supergravity probe limit (L2)}
\label{sec:l2}
In the previous section we found that the terms with $L = m$ and $n=0$
in the gauge theory effective action were associated with
supergravity.   In this section we will take another limit where
back-reaction of the probe is suppressed so that the supergravity
effective action in Eq.~\ref{eq:grav} can be directly compared to the
gauge theory action.  The new limit is defined by the substitutions:
\begin{equation}
L2:~~~~~~~
R \rightarrow \lambda R_0 ~~~~~;~~~~~
N \rightarrow \lambda^2 N_0 ~~~~~;~~~~~
K \rightarrow \lambda K_0
\label{eq:l2limit} 
\end{equation}
followed by $\lambda \rightarrow\infty $. Again, $R_S \rightarrow 0$ before
these limits are taken.  Following the analysis of the previous section,
the longitudinal momentum
density of the target $P_t^+/R \sim N/R^2$ is fixed in the $L2$
decompactification limit.  However, the probe longitudinal momentum
density $P_p^+/R \sim K/R^2$ and the transverse momentum density
$P_\perp/R \sim Kv^2$ vanish as $\lambda \rightarrow \infty$ although
the total probe momenta $P_p^+ \sim K/R$ and $P_\perp \sim Kv^2/R$
remain finite.  We have already argued that the $L1$ limit should be a
nonsingular limit of M theory.  The same discussion applies here,
with the added statement that the vanishing energy density of the
probe implies that it can be treated as moving in the background
produced by the target without any back-reaction. 

Examining the effect of the $L2$ scaling limit on the gauge theory
effective action in Eq.~\ref{eq:general} shows that the tree level
term $Kv^2/R$ remains finite.  This is because, unlike the $L1$ limit,
the total energy of the probe remains finite.  The general term in the
effective action scales as $\lambda^{2(L - 2n -m) - q}$.   Again, as
with the $L1$ limit, we have good reasons to assume that the $L2$
limit is nonsingular and so all divergent terms must in fact have
vanishing coefficients.   The finite terms satisfy $ 2n + m = L - q/2$.
However, the $L1$ limit has already shown that terms with $2n + m < L$
must vanish.   So only the terms with $q=0$ survive the $L2$ limit.   
We are left with the following finite terms in the effective action:
\begin{equation}
L_{L2} = {K \, v^2 \over 2\, R} ~\left[1+
\sum_{L = 1}^\infty\, d_{LL}(0,0)~ 
\left({N\,v^2\over R^2 M^9_P V_T~b^{7-p}}\right)^{L}~(1+S_{L2})
\right]
\label{eq:L2action}
\end{equation}
Here $S_{L2}$ is a series of corrections suppressed by inverse powers
of $M_P^6$:
\begin{equation}
S_{L2} =\sum_{n=1}^\infty f_{L0}(n) \, 
\left({1 \over M_P^3 V_T~b^{3-p} }\right)^{2n}
\end{equation} 
The series $S_{L2}$ collects the contributions from nonplanar diagrams
that survive the $L2$ limit.  The terms in $S_{L2}$ are suppressed by
powers of $1/M_P^6$ which matches our expectations for the corrections
to the low-energy effective action of M theory in the decompactified
limit~\cite{rtR4}.  The $n=0$ terms arise from planar diagrams and
should be compared to supergravity.

Therefore we compare the planar ($n=0$) diagrams that survive the $L2$
limit to the supergravity probe effective action in Eq.~\ref{eq:grav}.
Evidently, the gauge theory loop expansion coincides up to the unknown
coefficients $d_{LL}(0,0)$ with the velocity expansion of the
supergravity effective Lagrangian.  In the limit of parameters we are
studying, as we have argued above, both the gauge theory and the 11
dimensional supergravity analyses appear to be valid.  So the probe
Lagrangian in Eq.~\ref{eq:grav} is expected to be the summation of
planar diagrams that survive the $L2$ limit.   In other words, we
conclude that the summation to all loops of the leading $N$ planar
diagrams in the gauge theory effective action gives a Born-Infeld
action of the form:
\begin{equation}
L = - {K\over R} D^{-1} (\sqrt{1 - D v^2} - 1)
\end{equation}
One way of stating this result is that the leading large $N$ terms in
the gauge theory effective action are controlled by a symmetry - 11
dimensional Lorentz invariance - that causes them to resum to a
Born-Infeld form.


\paragraph{Independent Evidence for Born-Infeld resummation:}

The same $U(N+K)$ gauge theories studied above can be derived by
keeping $R_S$ fixed and examining the short distance dynamics of
Dp-branes where the probes have energies $E \sim (1/2) K v^2 R_S/R^2$
with $R \gg R_S$.  It is well known that for any $p$ this low-energy,
short-distance dynamics is governed by $p+1$ dimensional gauge
theory~\cite{witten95a,dkps}.  The gauge theory should be understood
here as merely a low-energy effective description of the branes - in
previous sections, the $R_S \rightarrow 0$ limit made gauge theory the
exact theory for $p \leq 3$.

In this alternative scenario, the analysis of the gauge theory
effective action for scattering of $K$ probe from $N$ target branes
proceeds exactly as in Sec.~\ref{sec:DLCQ}.  As
before, we separate the probes spatially from the target by a distance
$\tilde{b}$, and we give the probes a velocity represented by the
electric field Eq.~\ref{eq:electric}.  In addition, we turn on
constant magnetic fluxes restricted to live in $U(1)^K \in U(K)$ in
order to avoid difficulties when the fluxes do not commute.  Then, for
any $p$, ignoring cutoff dependent terms, the effective action
governing scattering of $K$ probes from the target will be given by
Eq.~\ref{eq:general} with the following modifications.  Each
$v^{2m+2}$ is replaced by a Lorentz scalar constructed from $(2m+2)$
factors of $F$, the flux on the probe.  Also, each factor of $K$ is
replaced by a trace over the probe gauge fields.  When there is more
than one trace it is hard to know how to distribute them over the
factors of $F$ and the relative coefficients between the different
Lorentz scalars at $O(F^{2m+2})$ should be determined.  (These issues
have been discussed in~\cite{chepact,peract}.)

The effective actions for various $p$ all arise as low-energy limits
of open-string theory.  Since T-duality is a symmetry of open-string
theory,  we expect that effective actions for different $p$ should be
related to each other by dimensional reduction~\cite{chepact}.  We
will use this expectation along with supersymmetry and existing
calculations to derive the leading $N$ gauge theory effective
action for any $p$ up to 2-loops.   The tree level Lagrangian (after
integrating the constant fluxes over the Dp-brane worldvolume) is:
\begin{equation}
L_0 = {{\rm Tr}(F^2) \over 4 R }
\end{equation}
We are interested in the leading $N$ planar diagrams which, as
discussed in previous sections, have $n=0$ and are proportional to
$N^L$ at $L$ loops.  After integrating over the $p$ spatial
dimensions, the general form of the terms with $(2L+2)$ factors of $F$
at 1 and 2 loops is:
\begin{eqnarray}
L_1 &=& (1/R) {\rm Tr}( \alpha \, F^4 + \beta \, (F^2)^2)\\
L_2 &=& (1/R) {\rm Tr}( \gamma \, F^6 + \sigma \, F^4 \, F^2 +
                       \tau \, (F^2)^3)
\end{eqnarray}
(We use the convention that $F^{2n}$ has a cyclic trace on Lorentz
indices.)

   When $F$ is a pure electric field (transverse velocity) as in
Eq.~\ref{eq:electric}, the value of $L_1$ and $L_2$ have been computed
in~\cite{dkps,bbpt}.  When $p=4$ and $F$ is self-dual, supersymmetry
is unbroken and there is no force between probe and target.  So in
this case $L_1$ and $L_2$ vanish.  Finally, if $p=4$, and we turn on a
self-dual magnetic flux as well as the transverse velocity in
Eq.~\ref{eq:electric}, we expect a metric on moduli space that is
exact at one loop~\cite{dkps}.  This gives another vanishing condition
on $L_2$.

Altogether we have five equations for five unknowns which can be solved
to give:\footnote{The identity $F^{2n} = (1/4^{2n-1}) (F^2)^n$ for
self-dual fields is very useful in showing this.}
\begin{eqnarray}
L_1 &=& {D \over R} \, {\rm Tr}\left( {F^4 \over 8} - {(F^2)^2 \over 32}
\right)  \\
L_2 &=& {D^2 \over R} \, {\rm Tr}\left( {F^6 \over 12} - {F^4 \, F^2 \over 32} +
        {(F^2)^3  \over 384} \right)
\end{eqnarray}
Here $D$ is the harmonic function in Eq.~\ref{eq:boostharm}.  With
some work it can be shown that $L_0 + L_1 + L_2$ is the expansion to
$O(F^6)$ of the Born-Infeld like Lagrangian:
\begin{equation}
L_{BI} = -(1/R) \,D^{-1} \, {\rm Tr}\left(\sqrt{\det(\eta_{\mu\nu} I - D^{1/2}
F_{\mu\nu})} - I\right)
\label{eq:BI}
\end{equation}
The determinant is over the Lorentz indices and $I$ is the $U(K)$
identity.  It is rather non-trivial that the one and two loop
effective Lagrangians fit the expansion of the Born-Infeld formula.
Previously, when $F$ simply represented a transverse
velocity, we derived the same Born-Infeld form for the leading $N$
planar diagrams that survive the $L2$ limit.  (In that case,
$\det(\eta_{\mu\nu} I - D^{1/2} F_{\mu\nu}) = (1 - D v^2) I$.) That
derivation used 11 dimensional Lorentz invariance and used the
plausible existence of  certain decompactification limits of M theory.
Here we have made the rather different assumption that gauge theory
effective actions are related by T-duality.  As a consequence of this
assumption, supersymmetry and existing calculations provide
significant evidence for the Born-Infeld resummation of the leading
$N$ planar diagrams studied in previous $L2$ limit.  This can 
be seen as partial evidence for the existence of the
decompactification limit assumed in previous sections.

\subsection{Another large $N$ decompactification limit (L3)}
\label{sec:twolim}

In this section we will analyze a large $N$ limit of a different sort from
before. Let us consider nearly null compactified M-theory specified by $N,
R$ and $\epsilon={R_s \over R}$ , a small parameter measuring the almost
``light-like-ness''. We will assume that if a decompactified limit
\begin{equation}
N, R \rightarrow \infty , \epsilon \rightarrow 0
\label{eq:l3limit}
\end{equation}
exists and is well defined, it is unique. This is a rather strong
assumption, but it seems reasonable in that if there are a number of
consistent limits then there would seem to be many noncompact
11-dimensional vacua.  So let us make this assumption and see what
conclusions we are led to. We will then see some evidence for it.

In particular, let us examine two separate ways of taking the limit in
Eq.~\ref{eq:l3limit}.
\begin{figure}                                 
\begin{center}                                 
\leavevmode                                 
\epsfxsize=3.5in
\epsfysize=2.5in                                 
\epsffile{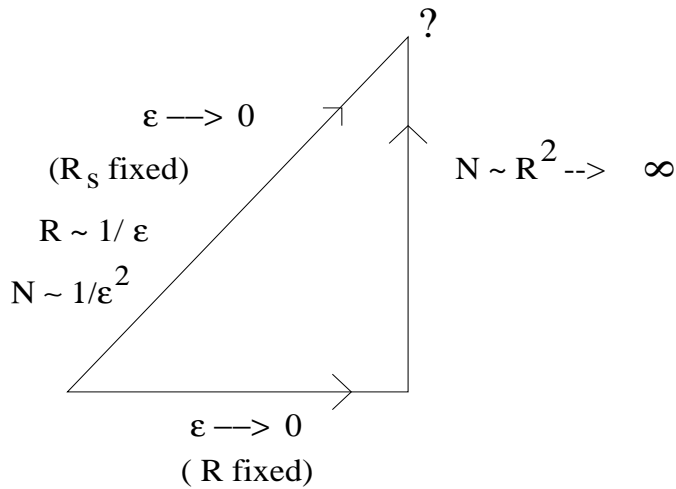}                                 
\end{center}                                 
\caption{Two possible decompactification limits}
\label{fig:triang}
\end{figure}
The first limit is the one that we have been studying thus far.  We start
in the lower left corner of the triangle in Fig.~\ref{fig:triang}, at some
given values of $N$, $R$ and $\epsilon$.  We then follow the horizontal
line in figure by taking the $\epsilon \rightarrow 0$ limit following
Seiberg~\cite{seibmat}.  In this limit $N$ and $R$ are finite.  As shown by
Seiberg, this limit of M-theory, at the lower right corner of the figure,
is related by a boost to weakly coupled strings in the gauge theory
limit. The vertical line then takes $N, R \rightarrow \infty$ following the
earlier L1 or L2 limits which, we argued, led to decompactified M theory.
Thus, finally, the $N=\infty$ gauge theory describes decompactified
11-dimensional physics.

The second limit that we can take is the one sketched along the diagonal of
the figure.  Here we simultaneously take $N, R \rightarrow \infty $ and
$\epsilon \rightarrow 0$ by keeping $R_s$ fixed.  We will refer to this as
the ``diagonal limit''.  As before, the precise way in which we take $N
\rightarrow \infty$ depends on what we want to study. In particular one
could consider either of the L1 or L2 limits.  After repeating Seiberg's
boost and rescaling of units as explained in Sec.~\ref{sec:DLCQ}, we find
ourselves in string theory.  This time we are in a regime where
$\tilde{g}_S = (R_S M_P)^{3/2}$ is fixed and $\tilde{M}_S^2 = {R^2 M_P^3
\over R_S} = {R_S M_P^3 \over \epsilon^2} $.  The velocities are $\tilde{v}
= v\epsilon$ and the transverse separations $\tilde{b} = b\epsilon$. Unlike the
previous $R_S \rightarrow 0$ limit, the physical separations in string
units $\tilde{b}\tilde{M}_S = b(R_S M_P^3)^{1/2}= bM_P \tilde{g}_S^{1/3}$
are fixed.  So we are not in the gauge theory limit of string theory - the
limit along the diagonal in Fig.~\ref{fig:triang} is related to a rather
different limit of string theory.  If this also describes decompactified
M-theory, then an equality of limits would imply that these two limits of
{\it string theory} are the same.  At first sight this seems completely
unlikely and therefore to undercut our initial assumption. But let us
examine it a little more.

We work in the domain where the decompactified theory would be adequately
descibed by 11-dimensional supergravity which we can then compare with our
second limit of string theory.  To this end we take $bM_P \gg 1$ so that
for fixed (small) $\tilde{g}_S$ we have $\tilde{b}\tilde{M}_S \gg 1$.  This
takes us to a domain the string theory is in a 10-dimensional supergravity
limit.  For concreteness, consider the process of two graviton scattering
in the probe limit (L2) Eq.~\ref{eq:l2limit}. In string theory we are then
computing the effective action for the probe D-particle in the presence of
the background fields produced by the target D-particle. For small string
coupling this is given by the disc with Dirichlet boundary conditions in
the presence of background fields.
\begin{eqnarray}
S_{\rm eff}&=& {K\tilde{M}_S \over \tilde{g}_S}\int dt\{e^{-\phi}(detG)^{1/2}-A
\}\nonumber \\ 
&=&{K\tilde{M}_S \over \tilde{g}_S}\int dt f^{-1}(\sqrt{1-f\tilde{v}^2}-1)
\label{eq:disc}
\end{eqnarray}
where 
\begin{equation}
f=1+{c\tilde{g}_S N\over (\tilde{r}\tilde{M}_S)^7}
\label{eq:f}
\end{equation}
As we take the limit $\epsilon\rightarrow 0$ (keeping $R_s$ fixed in the
``diagonal limit'' in the figure),  we see that the disc effective action
has a well defined limit:
\begin{eqnarray}
S_{\rm eff}&=& - \int dt {K\over R} D^{-1} (\sqrt{1 - D v^2} - 1) \\ 
D &=&  { c_0 \ N \over (r M_P)^{7} (R M_P)^2}
\end{eqnarray}
Note that the leading $1$ in Eq.~\ref{eq:f} has dropped away in the limit.
We are left with precisely the same action as the {\it 11-dimensional}
lightcone supergravity effective action that we had in
Eq.~\ref{eq:grav}. This is what we had earlier claimed would be the leading
term to survive in the gauge theory limit of string theory in the L2
limit. This is then some evidence that our conjecture of equality of
different limits of string theory is not altogether meaningless.  Note that
all dependence on $R_S$, which was fixed, has gone away.  This is
essentially like being in the near horizon regime of black holes in string
theory. The simultaneous validity of large $N$ gauge theory and
supergravity in this domain has already been proposed~\cite{dps}. Our
argument about the existence and equality of various decompactification
limits is a sort of a generalisation of this to general string theory
processes.

We can make further testable statements. In the second string limit, there
are corrections beyond the disc diagram in a background field. These are of
two kinds. There are the diagrams with one boundary and any number of
handles attached. Therefore they come with only one power of $K$ and hence
are corrections to supergravity but nevertheless in the probe limit. There
are corresponding candidate $M_P^6$ corrections in the gauge theory from
below the diagonal. It will be interesting to verify that, say, the disc
with one handle attached, in the presence of the target fields, gives the
same contribution as non-planar gauge theory diagrams from one below the
diagonal. If we go beyond the probe approximation one can consider string
diagrams with extra boundaries. Since these have higher powers of $K$, they
take into account the effect of the probe on the geometry.  Indeed, our
conjecture leads us to expect that the cylinder diagram with both
boundaries on the probe, in the background produced by the source should
reproduce the gauge theory planar diagrams that are of order $K^2$ and
survive the L1 limit.

\section{Discussion}
\label{sec:conj}
\label{sec:problems}

Let us review the results we have obtained.  We considered M theory on
a nearly lightlike circle and a transverse torus with $N$ units of
longitudinal momentum.  We defined three kinds of large $N$ limits of
this system, and argued that they were nonsingular.  Furthermore, we
argued that 11 dimensional supergravity is valid in these limits
because the effect of the momentum in the supergravity solution is to
make the circle spacelike and big.  The M theoretic description of
graviton scattering then allowed us to infer properties of gauge
theory such as an infinite set of non-renormalisation theorems. The
compatibility and existence of the third of our limits led to a
conjectured relation within string theory between two very different
regimes.  This had surprising, but testable, consequences.

The large $N$ limits that we have taken are not conventional 't Hooft
limits. In fact, in ordinary gauge theories, if we take the large $N$
limit keeping the gauge coupling finite, the Feynman diagram expansion
is wildly divergent. Why then do we expect our limits to be well
defined?  Our argument has been indirect, relying on the well-behaved
nature of supergravity.  We can turn this around and say that if
supersymmetric Yang-Mills has vanishing coefficents for the infinite
class of terms that we identified, then the unconventional large $N$
limits we defined are finite.

The argument that supergravity is valid at short distances in the large $N$
limit is also an important ingredient in other recent
works~\cite{near5,juanN}. In the latter reference it is argued that
superconformal invariance causes the leading $N$ terms in the effective
action of $3+1$ dimensional Yang-Mills to be of the Born-Infeld form. 
 It would be interesting
to understand the relation between our arguments and those of~\cite{juanN}.

Several recent works~\cite{issues,ggr1, dharetal, 
mathard,multigrav,pershort} have
reported disagreements between finite $N$ calculations in gauge theory and
supergravity. There is no contradiction between these results and ours
because we are dealing with the large N limit. 
We will, however, comment briefly on the reported finite $N$ discrepancy 
in the simplest generalisation of our setting - 
three graviton scattering~\cite{multigrav}.
Consider two gravitons at a separation of
$r$ and another, much further away at a distance $R$.  According to
perturbative supergravity in 11 dimensions, the interaction includes a term
of the form:
\begin{equation}
L_{3g}\sim \kappa^4~N_1 N_2 N_3 
{(v_1 -v_2)^2 (v_2 - v_3)^2 (v_1 - v_3)^2 \over r^7 R^7}
\label{eq:dineraj}
\end{equation}
It was argued in~\cite{multigrav} that such a term cannot arise in finite
$N$ gauge theory.  We have interpreted this scattering problem in 10
dimensions as the interaction of 3 D-particles that are almost at rest. In
the gauge theory the relevant problem involves the symmetry breaking
$SU(3)\rightarrow U(1)^2$. This can be analyzed in two parts, by first
considering $SU(3)\rightarrow U(2)$, which leads to an effective action
with vertices of the $v^4$ form.  This modified effective $U(2)$ Lagrangian
has mildly broken supersymmetry, so that the subsequent step
$U(2)\rightarrow U(1)^2$ gives rise to a $v^2$ force instead of $v^4$. In
this way we do arrive at terms of the general form $\kappa^4 \, {N_1 N_2
N_3 v^6 \over r^7 R^7}$.  The remaining issue is whether we reproduce the
precise kinematic factor in Eq.~\ref{eq:dineraj}.

The crucial second step that gives a $v^2$ force can be interpreted in
supergravity as the interaction between a D-particle and a near-extremal
D-particle~\cite{juanprobe}. We have exploited this to establish a
quantitative agreement between supergravity and gauge theory -- a specific
contribution to the 3 D-particle interaction.  This contribution from gauge
theory already serves to indicate the source of the tension between
supergravity and M(atrix) theory in~\cite{multigrav}: the gauge theory does
not exhibit manifest Lorentz invariance in 11 dimensions. There {\it are}
contributions of roughly the right form, but they do not naturally organize
themselves into Lorentz invariants, and so all terms must be calculated
before we can compare with Eq.~\ref{eq:dineraj}.\footnote{Work in
progress.}

The authors of the various papers pointing out difficulties in
matching gravity from the M(atrix) model have suggested that the problems
may disappear in the large $N$ limit.  In this paper we have defined
several several large $N$ limits that appear to have sensible low energy
supergravity descriptions in M theory.  As such, this is the arena in which
the M(atrix) model should confront supergravity.

\vspace{0.2in} 

{\bf Acknowledgments:} We would like to thank Shanta de Alwis, Shyamoli
Chaudhuri, Mike Dine, Eric Gimon, David Kastor, Albion Lawrence, David
Lowe, Juan Maldacena, Samir Mathur, Djordje Minic, Hirosi Ooguri, Arvind
Rajaraman, Sanjaye Ramgoolam, Nati Seiberg, Andy Strominger and Jennie
Traschen for useful discussions.  V.B. thanks the University of
Massachusetts at Amherst for hospitality and facilities.  F.L. thanks MIT
for hospitality while this work was completed.  V.B. is supported by the
Harvard Society of Fellows and by the NSF grant NSF-Phy-91-18167.  R.G. is
supported in part by DOE grant DOE-91-ER40618.  F.L. is supported in part
by DOE grant FG02-95ER40893.


\end{document}